\documentclass[conference]{IEEEtran}
\IEEEoverridecommandlockouts
\usepackage{cite}
\usepackage{amsmath,amssymb,amsfonts}
\usepackage{algorithmic}
\usepackage{graphicx}
\usepackage{textcomp}
\usepackage{xcolor}
\usepackage{comment}
\usepackage{subcaption}
\usepackage{multirow}
\usepackage{booktabs}
\usepackage{url}
\usepackage{balance} 

\def\BibTeX{{\rm B\kern-.05em{\sc i\kern-.025em b}\kern-.08em
    T\kern-.1667em\lower.7ex\hbox{E}\kern-.125emX}}
\begin{document}

\title{The ``AI+R"-tree: An Instance-optimized R-tree\\
\thanks{To appear in the proceedings of The 23rd IEEE International Conference on Mobile Data Management (2022)}
}

\author{\IEEEauthorblockN{Abdullah-Al-Mamun}
\IEEEauthorblockA{\textit {Purdue University} \\
mamuna@purdue.edu}
\and
\IEEEauthorblockN{Ch. Md. Rakin Haider}
\IEEEauthorblockA{\textit{Purdue University} \\
chaider@purdue.edu}
\and
\IEEEauthorblockN{Jianguo Wang}
\IEEEauthorblockA{\textit{Purdue University} \\
csjgwang@purdue.edu}
\and
\IEEEauthorblockN{ Walid G. Aref}
\IEEEauthorblockA{\textit{Purdue University} \\
aref@purdue.edu}
}
\maketitle

\begin{abstract}
The emerging class of instance-optimized systems has shown potential to achieve high performance by specializing to a specific data and query workloads. 
Particularly, Machine Learning (ML) techniques have been 
applied successfully 
to build various instance-optimized components (e.g., learned indexes). 
This paper investigates to leverage ML  techniques to 
enhance the performance of spatial indexes, particularly the R-tree, for a given data and query workloads. 
As the areas covered by the R-tree index nodes overlap in space, upon searching for a specific point in space, multiple paths from root to leaf may potentially be explored. In the worst case, the entire R-tree could be searched.
In this paper, 
we define and use the overlap ratio to quantify the degree of extraneous leaf node accesses required by a range query. The goal is to enhance the query performance of a traditional R-tree for high-overlap range queries as they tend to incur long running-times. 
We introduce a new AI-tree that transforms the search operation of an R-tree into a 
multi-label classification 
task
to exclude the extraneous leaf node accesses. 
Then, we augment a traditional R-tree to the AI-tree to form a hybrid ``AI+R''-tree. The ``AI+R’’-tree can automatically differentiate between the high- and low-overlap queries 
using a learned model.
Thus, the ``AI+R''-tree processes high-overlap queries using the AI-tree, and the low-overlap queries using the R-tree. Experiments on real  datasets demonstrate that the ``AI+R''-tree can enhance the query performance over a traditional R-tree by up to 500\%. 
\end{abstract}



\section{Introduction}
Traditional spatial indexes have been used successfully over the years as an efficient access method for location data.
In the area of spatial databases, 
the R-tree~\cite{guttman1984r} is 
a widely 
used index structure. 
In the multi-dimensional space, the R-tree is analogous to the one-dimensional index structure B$^{+}$-tree~\cite{comer1979ubiquitous}. 
These traditional index structures, e.g., the B$^{+}$-tree or the R-tree, do not make any assumptions about the underlying data distribution.
They are 
designed to work on a variety of data and query 
workloads.
As a result, an index is not necessarily optimized for a particular data and query 
workloads. 

Recently, there is an emerging class of instance-optimized systems proposed to optimize system performance for a specific data and query workloads, e.g.,~\cite{kraskatowards,ding2021instance}. 
Following the same direction, we target to design an index for a particular data and query workloads, i.e., an instance-optimized index; a learned index that has better search and lower space requirements than their traditional counterparts~\cite{kraskatowards, kraska2018case,al2020tutorial}.
Particularly, ML techniques have been successfully applied to build instance-optimized system components~\cite{kraska2018case, ding2021instance}. Although ML models are normally trained to generalize over a variety of datasets, in the context of designing instance-optimized components, overfitting of ML models can be desired if the models learn only from a known dataset~\cite{kraskatowards}.

\begin{figure}[tbp] 
     \centering
     \begin{subfigure}[b]{0.28\textwidth}
         \centering
         \includegraphics[width=\textwidth]{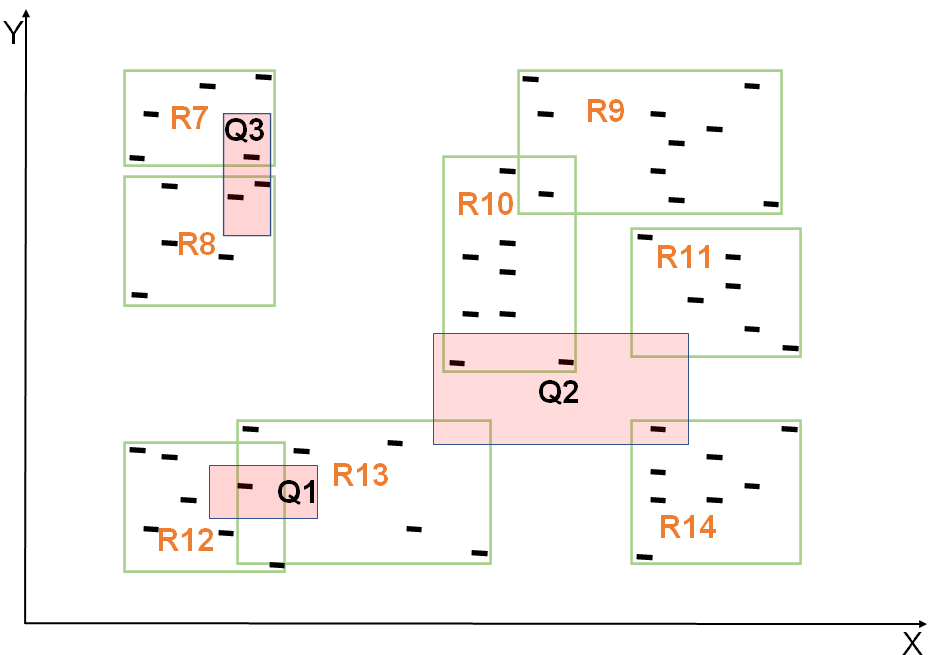}
         \caption{An example of leaf nodes in an \newline R-tree }
         \label{fig:leaf_nodes}
     \end{subfigure}
     \hfill
     \begin{subfigure}[b]{0.2\textwidth}
         \centering
         \includegraphics[width=\textwidth]{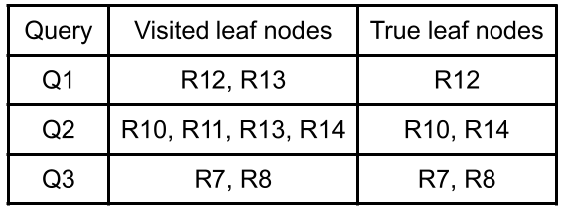}
          \caption{List of the accessed leaf nodes during query processing}
         \label{fig:range_queries}
     \end{subfigure}
     \hfill
        \caption{An example of R-tree range query processing} 
        \label{drawback_R-tree}
\end{figure}


In this paper, we focus on answering range and point queries over an R-tree due to their wide applicability in spatial databases~\cite{manolopoulos2010r}. 
In the R-tree, objects are stored using Minimum Bounding Rectangles (MBRs). Notice that in the B$^{+}$-tree, nodes do not overlap in space. However, the MBRs  of non-leaf and leaf nodes of an R-tree can overlap in space. 
Figure~\ref{drawback_R-tree} 
illustrates the impact of node 
overlap
in an R-tree to answer a range query. Only the MBRs of the leaf nodes are displayed in  Figure~\ref{drawback_R-tree}. 
Notice that the number of accessed leaf nodes directly impacts the query response time of an R-tree~\cite{manolopoulos2010r}. For a disk-based R-tree, descending multiple paths in the R-tree incurs high I/O cost~\cite{LISA}.
The leaf nodes of the R-tree 
are labelled R7-R14. 
Consider Range Queries Q1, Q2, and Q3 in Figure~\ref{drawback_R-tree}. To process Q1, the R-tree searches both R12 and R13, but the output data object is only present in R12. Hence, accessing R13 is wasted.
Similarly, to process Q2, the R-tree 
searches
R10, R11, R13 and R14, but the output data entries are only in R10 and R14. 
In both 
Q1 and Q2, the R-tree accesses $50\%$ more leaf nodes than the true number of leaf nodes containing the data objects. 
In contrast,
for Query Q3, the R-tree 
searches 
both R7 and R8, and data objects are exactly found in both 
nodes. 

In this case, the number of visited leaf nodes by the R-tree matches the true number of leaf nodes required to answer Q3. 
Thus,
in terms of the number of leaf node accesses, we can identify 
Q1 and Q2 
as
high-overlap queries  and Q3 
as
a low-overlap query. 
Observe that the R-tree searches extraneous leaf nodes to answer 
Q1 or Q2
but
performs optimally for Q3.
\begin{figure}[htbp]
  \centering
  \includegraphics[width=0.7\linewidth]{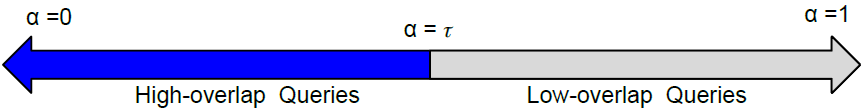}
  \caption{Spectrum of the overlap ratio $\alpha$ with Threshold $\tau$ to identify high- and low-overlap queries.}
  \label{figure_threshold}
\end{figure}
We define an overlap ratio $\alpha$ to quantify the degree of extraneous leaf node accesses required by a query. Specifically, for a range query, we divide the number of true leaf nodes by the total number of visited leaf nodes to 
estimate
$\alpha$, e.g.,
in Figure~\ref{drawback_R-tree}, to answer Q2, the number of visited leaf nodes is $4$ while the number of true leaf nodes is $2$ making 
$\alpha$
$= 0.50$. Similarly, for Q1 and Q3, $\alpha$ is $0.50$ and
$1$, respectively.
Notice that the number of true leaf nodes cannot exceed the number of visited leaf nodes. 
Hence,
$\alpha$ ranges from $[0 - 1]$.

For the purposes of 
this 
paper,
the high- and low-overlap queries are 
determined
as follows: 
Based on a pre-defined threshold $\tau$, queries with overlap ratio $\alpha \leq \tau$ (i.e., closer to $0$) are high-overlap while queries with $\alpha > \tau$ (i.e., closer to $1$) are low-overlap. 
The spectrum of the of the overlap ratio $\alpha$ with Threshold $\tau$ is shown in Figure~\ref{figure_threshold}.
{\bf \em To process  high-overlap queries, we propose to find the true R-tree leaf nodes 
using 
Multi-label Classification}; 
a supervised ML task, where an input object can be classified into one 
or multiple categories at once~\cite{herrera2016multilabel}. For example, 
classifying a research paper into 
a Systems, Theory, or ML paper is a multi-label classification task as a paper 
can be 
both a Systems and ML paper. 
Analogously, we can cast answering a range query over the R-tree, 
as a multi-label classification task, where the classes are the R-tree leaf nodes, and
we need 
to find these nodes that overlap the range query and that contain the output objects to the query. 

Motivated by the benefits of instance-optimized components (e.g., learned indexes)  and considering the issue of node overlap in the R-tree,
the following important questions arise: 
\textit{Which workloads degrade the performance of range query processing in a traditional R-tree?
Can we leverage ML techniques 
to make R-tree range query processing faster? 
}

\begin{figure}[htbp]
  \centering
  \includegraphics[width=0.8\linewidth]{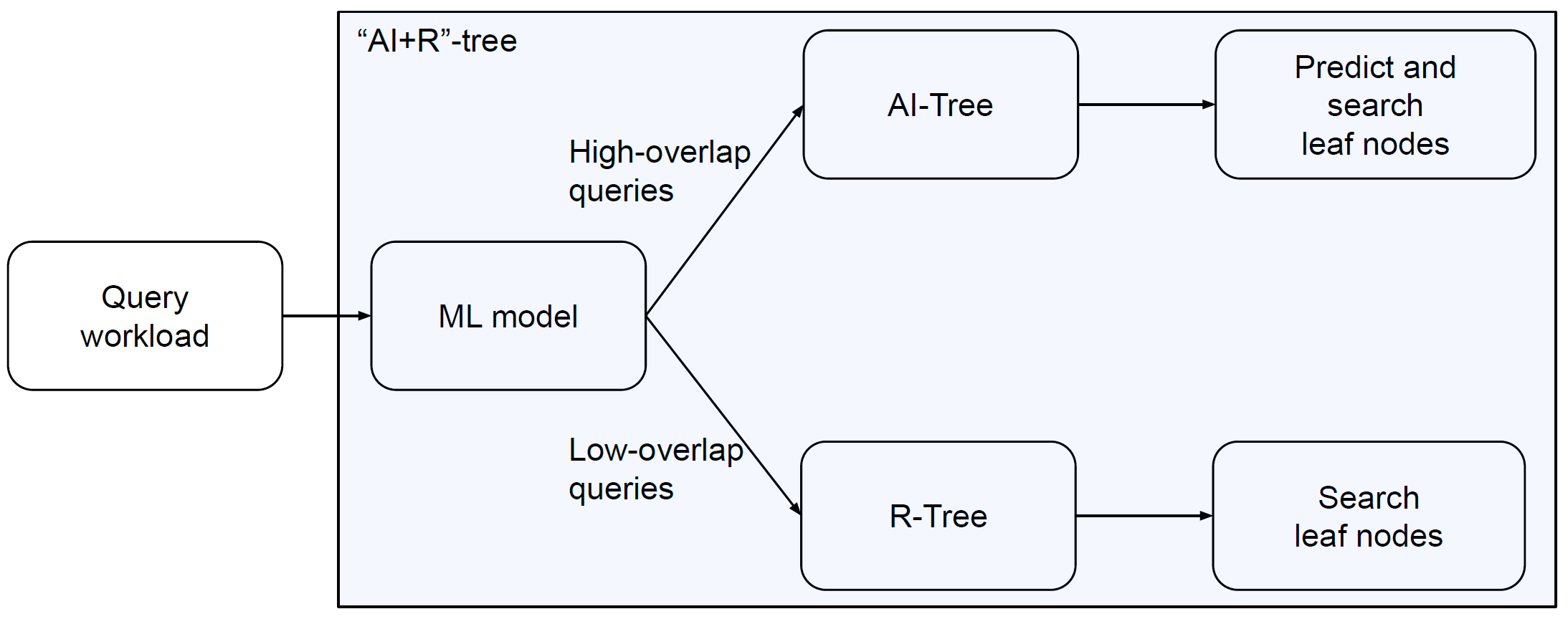}
  \caption{Workflow of the ``AI+R''-tree leveraging both the AI-tree and R-tree}
  \label{figure_hybrid_approach}
\end{figure}

We propose to use the 
overlap ratio $\alpha$ 
to 
identify the high-overlap queries for which an R-tree accesses many extraneous leaf nodes. Moreover, we propose to build an ML-enhanced R-tree, termed the AI-tree, 
that leverages 
multi-label classification techniques~\cite{herrera2016multilabel}.
Finally, we adopt a hybrid structure, termed the ``AI+R''-tree, to avail the benefits of both the AI-tree and the traditional R-tree 
(refer to Figure ~\ref{figure_hybrid_approach}).
The ideas behind the AI-tree is as follows: First, we perform a preprocessing step to assign IDs to the leaf nodes of the R-tree. 
Then, we treat the queries as input and the corresponding true leaf node IDs as class labels. In the example R-tree
in Figure~\ref{drawback_R-tree}, for $Q1, Q2,$ and $Q3$, the corresponding class labels are the IDs 
$\{R12\}$, $\{R10, R14\}$, 
and
$\{R7, R8\}$, 
respectively. 
Moreover, we 
prepare training data by processing each of the queries in a traditional R-tree and storing the corresponding true leaf node IDs as their class labels. Then, a multi-label classifier is constructed based on this training data. 
Motivated by the benefits of using multiple ML models (instead of a single model)~\cite{kraska2018case}, we adopt a multi-model approach that indexes the learned ML models in a grid-based structure.

To realize
the
``AI+R''-tree, we leverage a binary classification technique~\cite{aggarwal2015data_book} to learn the value of the overlap ratio $\alpha$ given an input range query.
This enables the ``AI+R''-tree
to differentiate between high- and low-overlap queries. 
Notice that the AI-tree is likely to perform better 
for
high-overlap queries while a traditional R-tree is expected to perform better 
for
low-overlap queries (due to the fact that there is limited scope for improvement).
Notice further that the AI-tree performs exact (i.e., not approximate) range query processing.
Thus, 
the ``AI+R''-tree leverages the benefits of both the AI-tree and the R-tree. 


The contributions of this paper are as follows:
\begin{enumerate}
  \item We introduce an instance-optimized AI-tree that transforms the R-tree search operation into a multi-label classification task. 
  While learned indexes are centered around the idea of {\em learning the index},
  the AI-tree adds to that the idea of {\em indexing the learned models}.
  \item We  leverage ML to  differentiate between high-
  and low-overlap range queries. This gives rise to the  ``AI+R''-tree that processes both query types efficiently.
  \item For fixed query workloads, experiments on real spatial data demonstrate that the
  ``AI+R''-tree enhances the performance of a traditional R-tree by up to 500\%.
\end{enumerate}

The remainder of this paper proceeds as follows: 
Section~\ref{problem formulation } presents the problem formulation.
Section~\ref{section:AI-tree} introduces the AI-tree. 
Section~\ref{section:``AI+R''-tree} introduces  the hybrid ``AI+R''-tree. 
Section~\ref{section:Evaluation} presents the experimental results.
Section~\ref{section:related_work} gives an overview of the related work.
Finally, Section~\ref{section:conclusion} presents concluding remarks and suggestions for future research.



\section{Background and Problem Formulation} \label{problem formulation }
\subsection{The R-tree: An Overview}
The R-tree~\cite{guttman1984r} is a balanced hierarchical index for multi-dimensional  objects. 
Each leaf or non-leaf node  of the R-tree  contains at least $m$ and at most $M$ entries. 
A rectangular range query is expressed as follows: Q ($X_{min}$, $Y_{min}$, $X_{max}$, $Y_{max}$), where
($X_{min}$, $Y_{min}$) and ($X_{max}$, $Y_{max}$) represent the bottom-left and top-right points of the query 
rectangle, respectively.
To process a range query $Q$, we 
start from the root of the tree, and check the MBR for each child of the root against Q to test which child nodes overlap  $Q$. In case of an overlap, we search the sub-tree rooted at the corresponding child. 
When we reach a leaf node, we   report the objects that overlap Q. 

\subsection{Classification: A Supervised Machine Learning Technique } \label{classification}
Classification~\cite{aggarwal2015data_book} is a commonly used ML technique.  
It can be divided into the following three categories: 
{\em (1) Binary Classification,} \label{sec_binary_classification}
where the number of classes is restricted to two,
{\em (2) Multi-class Classification,} where the number of classes $n$ is more than 2, and the goal is to classify an object into exactly one of the $n$ classes, and 
{\em (3) Multi-label Classification} \label{Multi-label Classification}
\cite{herrera2016multilabel,gibaja2015tutorial}, where we also have $n$($>$2) classes, but the goal is to classify an object into $c$ classes, where $0 \le c \le n$. 

\subsection{Problem Formulation}
  \begin{figure}[t!]
  \centering
  \includegraphics[width=\linewidth]{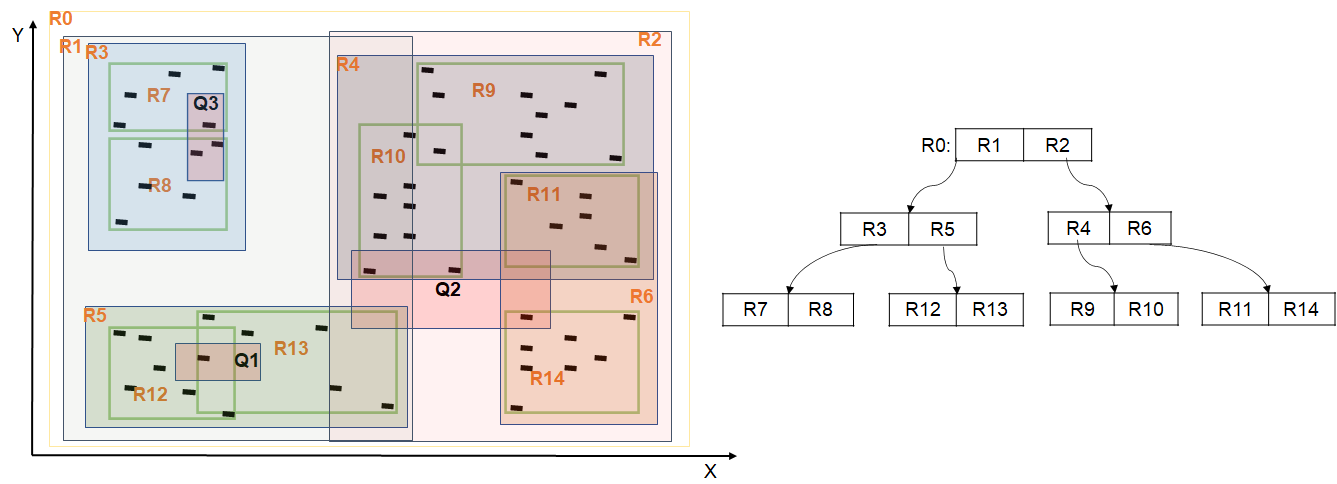}
  \caption{An example of R-tree with overlapping nodes}
  \label{node_overlapping}
\end{figure}
Refer to Figure \ref{node_overlapping} for illustration. 
Consider Queries Q1-Q3. For Q1, we search the R-tree down
the
2 paths (from root to leaf): $R1\rightarrow R5 \rightarrow R12$ and $R1 \rightarrow R5 \rightarrow R13$ while only 
the latter
path contains 
actual query results.
For Q2, we search the R-tree along 4 paths: $R1\rightarrow R5\rightarrow R13$, $R2\rightarrow R4\rightarrow R10$, $R2\rightarrow R6\rightarrow R11$, and $R2\rightarrow R6\rightarrow R14$ while only 
the 2nd and 4th paths contain output data objects. Notice that, for Query Q3, the R-tree searches two paths: $R1\rightarrow R3\rightarrow R7$ and $R1\rightarrow R3\rightarrow R8$, where both of them will contain output data objects.
Thus, for processing Q1 and Q2, the R-tree searches extraneous leaf nodes.
Hence, we formulate
our problem 
as follows: 
\textit{\bf Given a range query $Q(X_{min}, Y_{min}, X_{max}, Y_{max})$, we need to predict the true leaf nodes of the R-tree that contain 
output data objects, and only access these  nodes without accessing  extraneous ones.}

We propose to 
formulate this problem as a multi-label classification task. 
For example, assume that 
an R-tree has four leaf nodes
with 
unique IDs
1--4. For a range query, the R-tree may have to access any number of leaf nodes out of these four leaf nodes. We 
transform this problem into a multi-label classification task by treating the leaf node IDs as the class labels. 
At query time, the trained multi-label classifier 
predicts
the true leaf node IDs 
that
contain data entries that fall inside the query region. Hence, we only need to access the predicted leaf nodes to process the query. 


\section{The AI-tree } \label{section:AI-tree}


\subsection{The Preprocessing Phase}

\subsubsection{Assigning Unique Identifiers to the R-tree Leaf Nodes}
In the preprocessing step, each R-tree node is assigned a unique integer identifier (ID) based on a Depth First Search (DFS) order. Thus, all sibling leaf nodes of the
R-tree will have  consecutive integers as their IDs. 

\subsubsection{Definition of the Overlap Ratio $\alpha$}


We define an overlap ratio $\alpha$ to quantify the degree of extraneous leaf node accesses required by a range query. Given a range query Q, to calculate the value of $\alpha$, we use two metrics: the {\underline t}rue {\underline n}umber of leaf-node accesses required to process Q (TN(Q), for short), and the {\underline n}umber of leaf nodes {\underline v}isited by the R-tree index to answer Q (VN(Q), for short). 
For the range query Q, the definition of $\alpha$ is as follows (the value of $\alpha$ is in the interval 
 $[0,1]$):
\[ \alpha = \frac{TN(Q)}{VN(Q)} \]


\subsubsection{Query Workload Categorization} \label{Query Workload Generation}
Given a query workload, we categorize each query based on its selectivity.
After identifying the selectivity of a query,
the overlap ratio $\alpha$ of the query is calculated to further categorize the queries based on their value of $\alpha$. This is achieved by  executing the query during the preprocessing phase, computing the query's selectivity, the leaf nodes being touched, and the true leaf nodes.

\subsubsection{Preparing Training Data} \label{training_data_generation}
 This is a two-step process. In the first step, all queries in the query workload are executed one at a time on the constructed R-tree over the given dataset. For each executed query, we collect the following information: The IDs of the leaf nodes that the R-tree visits to answer the query, and the true leaf node IDs that contain the output data objects that are actually inside the query region. 

\begin{table}[h!]
\caption{Step-1 of Training Data Preparation} 
\centering 
\begin{tabular}{p{0.3\linewidth}p{0.3\linewidth}p{0.25\linewidth}} 
\toprule
Query&Visited Nodes&True Nodes\\
\midrule
$Q1$ & R12,R13 & R13\\
$Q2$ & R13,R10,R11,R14 & R10,R14\\
$Q3$ & R7,R8 & R7,R8\\
\bottomrule
\end{tabular}
\label{step1_training_data}
\end{table}

Assume that the ID assignment 
for the R-tree presented in Figure~\ref{node_overlapping} is as follows: R7 and R8 have IDs 1 and 2, R12 and R13 have IDs 3 and 4, R9 and R10 have IDs 5 and 6, and R11 and R14 have IDs 7 and 8. For Query Q1, the visited leaf nodes are R12 and R13 but the true leaf node is R13. 
Thus, for training purposes, for Q1, we set the ID of R13, i.e., 3, as the output label for the multi-label classifier problem. Similarly, for 
Query Q2, we have the ID of R10 and R14 (6 and 8) as the labels for the multi-label classifier.
Moreover, for Q3, we have the ID of R7 and R8 (1 and 2) as the labels. These steps are summarized in Tables~\ref{step1_training_data} and~\ref{step2_training_data}. In Table~\ref{step1_training_data}, for each query, we list the visited and the true leaf nodes. 
In Table~\ref{step2_training_data}, we list the leaf node IDs as the class labels for each of the queries.

\begin{table}[h!]
\caption{Step-2 of Training Data Preparation} 
\centering 
\begin{tabular}{p{0.1\linewidth}p{0.4\linewidth}p{0.2\linewidth}} 
\toprule
Query&Input Feature&Labels\\
\midrule
$Q1$ & $(X_{min},Y_{min},X_{max},Y_{max})$ & 3\\
$Q2$ & $(X_{min},Y_{min},X_{max},Y_{max})$ & 6, 8\\
$Q3$ & $(X_{min},Y_{min},X_{max},Y_{max})$ & 1, 2\\
\bottomrule
\end{tabular}
\label{step2_training_data}
\end{table}


\subsubsection{Feature Representation}~\label{freature_representation}
For an input range query Q, 
we use the values $(X_{min},Y_{min},X_{max},Y_{max})$ of the query rectangle as input features to the ML model without any additional transformation. Thus, the same input can be processed seamlessly by both the AI-tree and the R-tree. Moreover, for multi-label classification, the output labels are encoded using one-hot encoding, where we  represent the class labels using binary values, which is suitable for training the multi-label classifier, e.g.,
in Table~\ref{step2_training_data}, for query $Q1, Q2$ and $Q3$, the labels will be encoded as $00100000, 00000101$ and $11000000$.

\subsection{Learning the R-tree Index: ML Model Training and Testing}
\begin{figure}[h!]
  \centering
  \includegraphics[width=0.8\linewidth]{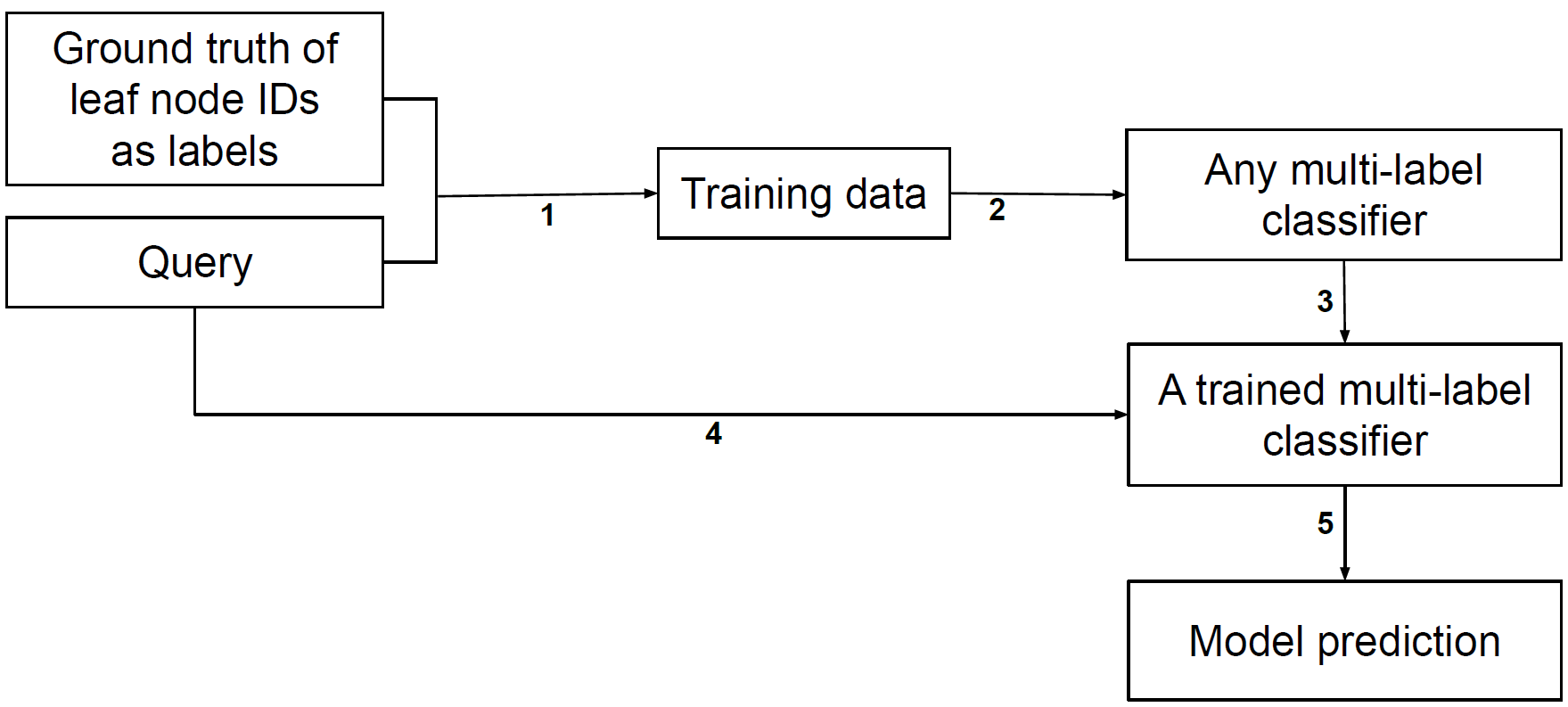}
  \caption{Workflow of model training and testing}
  \label{figure_workflow_model_training}
\end{figure}

Refer to Figure~\ref{figure_workflow_model_training}. The workflow for training and testing the multi-label classifier is as follows.
(1)~ While the given queries are executed by the R-tree, for each query, the IDs of the visited and true leaf nodes are captured. Thus, following the approach in Section~\ref{training_data_generation} (i.e., using the feature representation of the queries and the true leaf node IDs as labels),
the training data is prepared for a particular data and query workloads.
(2)~Then, the 
training data is used to train the multi-label classifier. Because the goal is to enhance the performance of the R-tree search operation for a fixed data and query workloads, the ML models are intentionally overfitted on the training data.
%
In this paper, we use a 
multi-label decision tree classifier~\cite{gibaja2015tutorial,quinlan1986induction} due to its ability to overfit the training data. 
Notice that the choice of the multi-label classifier is not limited to the family of decision tree classifiers. 
Because  R-tree search 
has been cast as a multi-label classification problem, 
we have the opportunity to use any suitable multi-label classifier~\cite{szymanski2019scikit_multilearn}.
(3)~A trained multi-label classifier will be created after the training phase.
(4)~As the AI-tree is optimized for a fixed query workload, the queries will be re-used as input for both the training phase of the  multi-label classifier and the testing phase. This approach is similar to the previous works that leverage overfitting to build instance-optimized systems components~\cite{kraska2018case,kraskatowards}.
(5)~At query time, the pre-trained multi-label classifier is invoked to directly predict the true leaf node IDs that contain the query result.

\subsection*{Indexing the Learned Models: A Multi-model Approach} \label{indexing the learned model}
The idea of indexing multiple learned models using a traditional index structure 
has been used 
in the context of music retrieval~\cite{jin2002indexing} and in 
 handwritten and time series data~\cite{handwritten_trie}. Moreover, the benefit of indexing the learned models using a recursive model index is also shown in~\cite{kraska2018case}.
Notice that for exact range query processing, our goal 
is to perfectly (i.e., 100\% prediction accuracy) fit the ML models to a particular data and query workloads.
However, even with overfitting, it might not be possible to train a single ML model to capture the entire underlying distribution of the training data~\cite{kraska2018case}. 
 As a result, to achieve high prediction accuracy on the 
 training dataset, multiple ML models 
 are trained, e.g., several multi-label decision tree classifiers instead of a single ML model. 
 
\begin{figure}[htbp]
  \centering
  \includegraphics[width=0.7\linewidth]{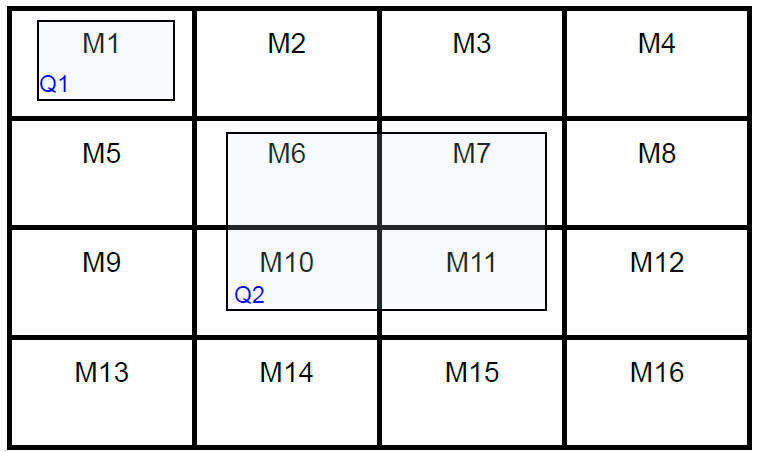}
  \caption{Indexing the learned models} 
  \label{multi-model_approach}
\end{figure}

 In the AI-tree, we use a simple index structure, e.g., a coarse grid to partition the training queries. Then, we train a separate ML model over 
 queries inside each grid partition. The grid serves as an index to the localized learned ML models. 
 As a result, at query time, we only invoke the ML models whose grid cell overlap the query rectangle. This concept is illustrated in Figure~\ref{multi-model_approach}. 
 The steps are as follows: Initially, the underlying space is divided into equally sized grid cells.
 Then, during the training phase, if a query overlaps 
 a single grid cell, 
 the ML model  corresponding to that cell is trained for that query. Similarly, if a query overlaps multiple grid cells, all the cells' corresponding ML models are trained for that query. Notice that if no query overlaps 
 a particular grid cell, we do not train any ML model for that grid cell. For example, for a $10X10$ grid, it is not always the case that $100$ ML models need to be trained. 
 Finally, during query processing, only the ML models whose grid cells overlap the range query are executed.

For example, in Figure~\ref{multi-model_approach}, the space is partitioned using a $4X4$ grid. Thus, at most 16 ML models (M1 to M16) can be trained using the grid. 
In the training phase, we incrementally search for the grid size 
that produces the best fit over the training data~\cite{bergstra2011algorithms}.
In Figure~\ref{multi-model_approach}, as Query Q1 falls completely inside the top left grid cell, Model M1 is trained for  Q1. On the other hand, as Query Q2 overlaps 
four grid cells, the four models M6, M7, M10, and M11 are trained for Q2. 
If multiple models are trained for a particular query, during query processing, we aggregate their prediction results from all the overlapping cells. This aggregated result is produced by performing a union of the predictions of the individual ML models. 

\subsection{Query Processing}

Given a range query, first, the ML models are identified whose grid cells overlap the range query. Then, only the designated ML models are executed to process the query. After the prediction of the ML models, the results are aggregated in terms of leaf node IDs. 
Then, only the  leaf nodes whose IDs have been predicted by the ML models are accessed. Finally, all the data entries inside these leaf nodes are scanned to check which entries are actually contained within the input query rectangle. 
This ensures that 
the 
AI-tree never produces a false-positive result. {\bf Notice that we only access the predicted leaf nodes without traversing the R-tree or accessing the non-leaf nodes or the extraneous leaf nodes. Thus, if we  predict the leaf nodes accurately, we  access the minimum number of leaf nodes needed to answer a query. This reduces the number of disk I/Os for processing a range query}. 

Notice that in rare cases, the multi-label classifier (Section~\ref{classification}) might not predict any label for a particular query. In other words, the classifier might predict no leaf node ID for a particular query. 
For the AI-tree, if the set of predicted leaf node IDs is empty for a particular query, 
we 
invoke a regular R-tree search operation.
Moreover, if an ML model predicts a leaf node that does not contain any data object
that 
is qualified in the result (i.e., mispredicts) of the given range query, we may resolve to search the regular R-tree. 
Thus, the AI-tree performs exact query processing by combining both the multi-model approach and the regular R-tree.

\section{The "AI+R"-tree } \label{section:``AI+R''-tree}
To achieve the best of 
both the AI-tree and 
the R-tree, we adopt a hybrid approach 
that we term the ``AI+R''-tree (see Figure~\ref{figure_hybrid_approach}). We process the high-overlap queries using the AI-tree and the low-overlap queries using the traditional R-tree. However, this is non-trivial because the overlap ratio $\alpha$ of a query is unknown until we process the query. Hence, we  leverage ML techniques to learn how to distinguish between high- and low-overlap queries.
Specifically, the problem of classifying the range queries based on the value of $\alpha$ and the threshold $\tau$ can be formulated as a binary classification task~\ref{sec_binary_classification}. 
In order to 
prepare the training data for a particular dataset, we combine the 
queries for each of the $\alpha$ values. 
Then, 
we assign Label $0$ for the queries whose $\alpha$ value is less than or equal to the threshold $\tau$, and assign Label $1$ for the queries whose $\alpha$ value is greater than the threshold $\tau$. Next, a binary classifier is trained on the training data. Finally, we can use the trained binary classifier to classify an incoming range query into either a high- or a low-overlap query. 

\subsection{Range Query Processing in the ``AI+R''-tree} \label{a hybrid approach}
Given a range query Q, the binary classifier is invoked first (see Figure~\ref{figure_hybrid_approach}) to predict whether the incoming query Q is high- or low-overlap. If Q is classified as a high-overlap query, the AI-tree processes the query. Otherwise, the R-tree processes the query. Notice that query processing using the ```AI+R''-tree incurs a prediction cost before accessing the leaf nodes. Hence, the cost of query processing of the ``AI+R''-tree is: ML model prediction cost + I/O cost. Thus, we expect to get the benefit of the AI-tree for processing the high-overlap queries whose $\alpha$ value is closer to zero. On the other side of the spectrum of $\alpha$ (Figure~\ref{figure_threshold}), for the queries with $\alpha$ closer to one, the R-tree is expected to perform better than the AI-tree.

To demonstrate query processing in the ``AI+R''-tree, consider the three queries in Figure~\ref{drawback_R-tree}. For Queries Q1 and Q2, the overlap ratio $\alpha = 0.50$. If the ``AI+R''-tree can accurately predict the leaf nodes, $50\%$ less number of leaf nodes will be accessed to answer the query. Notice that we have room for improvement to process Q1 and Q2 using the AI-tree component of the ``AI+R''-tree. In contrast, for Q3, $\alpha = 1$. Thus, both the visited leaf nodes contain data entries that fall inside the query rectangle. Thus, it is not possible for the AI-tree to process the query using less leaf node accesses than the R-tree. Thus, we use the R-tree in this case.



\section{Evaluation}\label{section:Evaluation}

We  run all  experiments on an Ubuntu $18.04$  with Intel Xeon Platinum $8168$ ($2.70$GHz) and $3$TB of total available memory.


\subsection{Datasets}
We use two datasets from the UCR Spatio-Temporal Active Repository, namely UCR-STAR \cite{GVE+19}. Specifically, we use two real-world datasets with two-dimensional location data (in the form of longitude and latitude). The Tweets location dataset contains the locations of real tweets, and the other dataset contains the locations of Chicago crimes. Moreover, we have 
preprocessed
the datasets before using them for the experiments. First, we eliminate the duplicate and missing values from both datasets. 
For the Tweets location dataset, we create a processed dataset containing the first $2$ million tweet locations. On the other hand, after removing the duplicate values from the dataset of Chicago crimes locations, we get a processed dataset containing $872,127$ records.  

\subsection{Parameter Settings}

\subsubsection{R-tree Parameters}
All R-tree variants attempt to reduce the amount of node overlap. However, with dynamic updates, the shape of the R-tree deteriorates over time. 
Thus, we construct the R-tree using a one-at-a-time tuple
insertion method to replicate the scenario of a dynamic environment. When constructing the R-tree, we set the minimum leaf node size $m$ to $50\%$ of the maximum
leaf node size $M$. 
Another parameter of 
the
R-tree is the node-splitting algorithm. In the experiments, we use the linear node-splitting algorithm.  

\subsubsection{Query Selectivity and Values of $\alpha$}
For a particular dataset, to demonstrate the query performance for a particular value of $\alpha$, (at most) 1000 synthetic range queries are
used in the experiments with a fixed selectivity. 
For example, in the case of the Tweets location dataset, a range query with Selectivity $0.00001$ returns approximately $20$ objects, and a query with Selectivity $0.00005$ returns approximately $100$ objects. In the experiments, the selectivity varies between $0.00001$ and $0.00005$. 
We categorize the queries into five different values of $\alpha [0.1, 0.25, 0.5, 0.75, 1.0]$.
Thus, 
for each dataset, we 
use up to $5000$ queries with various values of $\alpha$. 

\subsubsection{The ''AI+R''-tree Parameters}
The ``AI+R''-tree has two parameters: The size of the grid (see Section~\ref{indexing the learned model}) and the choice of the threshold $\tau$ (see Figure~\ref{figure_threshold}). 
Similar to the idea of hyperparameter tuning~\cite{bergstra2011algorithms} for ML models, we start from a grid size $2X2$ and increase the size (e.g., $4X4$) 
to get the best fit for the training data.
In all the experiments, we have achieved the 
best fit over the training data
with a maximum grid size of $20$X$20$.
Notice that using the multi-model approach
and invoking the regular R-tree in case of a misprediction,
the AI-tree can 
achieve $100\%$ prediction accuracy over the training data.
As a result, both the AI-tree and the ``AI+R''-tree can perform exact (i.e., not approximate) range query processing.

On the other hand, for a query Q with $\alpha = 0.75$, the $\frac{TN(Q)}{VN(Q)}$ can be e.g., $\frac{15}{20}$. Thus, there is room for improvement unless $\alpha =1$. 
Thus, we set Threshold $\tau =0.75$. 
In other words, for an incoming range query, if $\alpha \leq 0.75$, it is identified as a high-overlap query. 
If $\alpha > 0.75$, it is considered low-overlap.  

\subsection{Choosing the ML Models}

\subsubsection{The Multi-label Classifier}
A decision-tree classifier~\cite{quinlan1986induction} has the ability to overfit the training data, and hence can achieve high prediction accuracy for the training dataset. Moreover, a decision-tree classifier is simple and explainable. As a result, we use a multi-label decision tree model as the multi-label classifier~\cite{gibaja2015tutorial}. However, with proper training, any multi-label classifier~\cite{szymanski2019scikit_multilearn} can be used in the ``AI+R''-tree. For the ML models, we use the standard scikit-learn python library~\cite{scikit-learn}. We use the default parameters for the decision tree classifier except the maximum-depth that is
set to $30$. This maximum-depth is set to a high value to allow the decision tree classifier to overfit the training data. 

\subsubsection{The Binary Classifier}
For binary classification, the goal is to train an ML model to classify an incoming query as high- or low-overlap. Notice that the goal is not to overfit but rather to generalize, and the same learned model will be able to classify high- vs. low-overlap queries across different query workloads.
We use a random forest classifier~\cite{breiman2001random} as the binary classifier. However, with proper training, any binary classifier can be used in the ``AI+R''-tree. The training process of the binary random forest classifier is as follows: For a particular data and query workloads, we combine the 
queries for each $\alpha [0.1, 0.25, 0.5, 0.75, 1.0]$.
For a particular dataset with  fixed selectivity queries, we will have up to $5000$ queries in total. 
For the binary classifier, we create the training data as follows: We assign Label  $0$ for queries where $\alpha \leq \tau$ (e.g.,  $\alpha \leq 0.75$), and Label $1$ for queries where $\alpha > \tau$ (e.g., $\alpha$ $>$ $0.75$). Moreover, we  split the training data where we use $80\%$ for training and $20\%$ for testing. We use the scikit-learn python library~\cite{scikit-learn}, and use the default scikit-learn settings for the random forest binary classifier. 
The prediction accuracy of the binary random forest classifier is around $80\%$ over all values of $\alpha$.


\subsection{Implementation and Measurements}
We realize the ``AI+R''-tree using an open-source python library for the R-tree available on Github~\footnote{\url{https://github.com/sergkr/rtreelib}}.
We integrate the ``AI+R''-tree inside the library and run the experiments using Python Version 3.6.9. 
On the other hand, for a disk-based R-tree index realized inside a practical system, in most of the cases, only the leaf nodes are stored in disk pages, and the internal nodes are kept in-memory. As a result, the performance of a query depends on both the CPU cost and the number of leaf node accesses. 
In the experiments, we assume that the required number of disk I/Os is equivalent to the number of leaf node accesses~\cite{mahmood2014indexing}. For a query, we measure the CPU time, and count the number of leaf node accesses. 
Then, we multiply the number of leaf node accesses by a standard disk I/O access time. Finally, we sum the CPU and disk I/O times to report the average query processing time (in milliseconds). 
In the experiments, we 
use a disk I/O access time of thirteen milliseconds~\cite{deng2011future}. 
This approach is similar to the experimental setup of a previous work~\cite{mahmood2014indexing}.

Also, we report the size of the R-tree and the size of ML models~\footnote{\url{https://docs.python.org/3.6/library/sys.html}}. The size of the ML models contains the summation of the sizes of both the multi-label and the binary classifiers. 

Notice that for a particular query workload with fixed selectivity, to demonstrate the performance for each value of $\alpha$, we run each experiment individually for each value of $\alpha$. This enables us to report the average query processing time and the size of the ML models for each value of $\alpha$.

\subsection{Experimental Results}
\begin{figure*}[tbph] 
     \centering
     \begin{subfigure}[b]{0.24\textwidth}
         \centering
        \includegraphics[width=\linewidth]{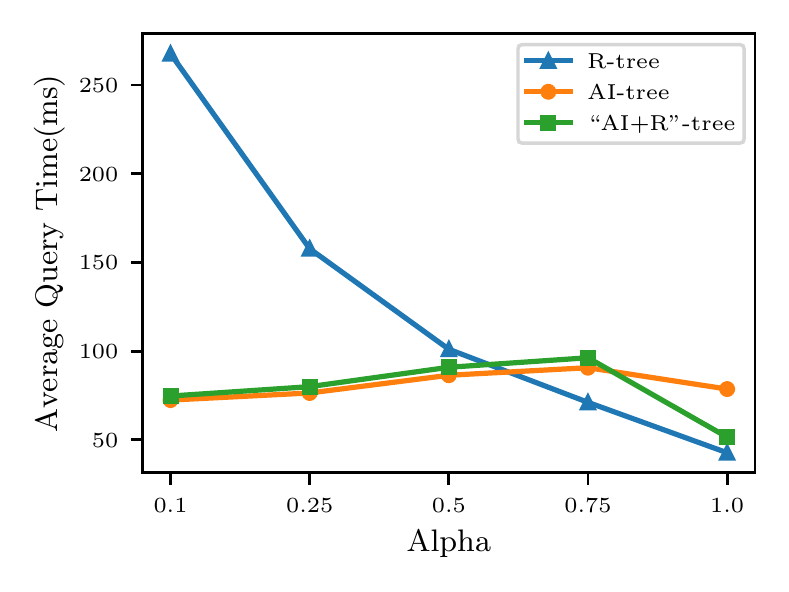}
        \caption{R-tree (M=200, m=100), and query selectivity = $0.00001$}
        \label{tw_ucr_me200_sel0.00001_20X20}
     \end{subfigure}
     \hfill
     \begin{subfigure}[b]{0.24\textwidth}
         \centering
         \includegraphics[width=\linewidth]{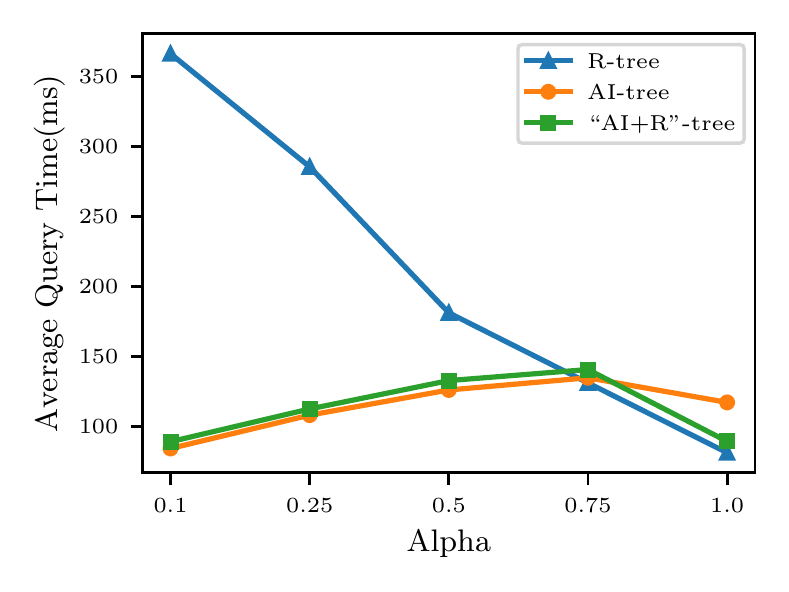}
        \caption{R-tree (M=200, m=100), and query selectivity = $0.00005$}
        \label{tw_ucr_me200_sel0.00005_20X20}
     \end{subfigure}
     \hfill
     \begin{subfigure}[b]{0.24\textwidth}
         \centering
         \includegraphics[width=\linewidth]{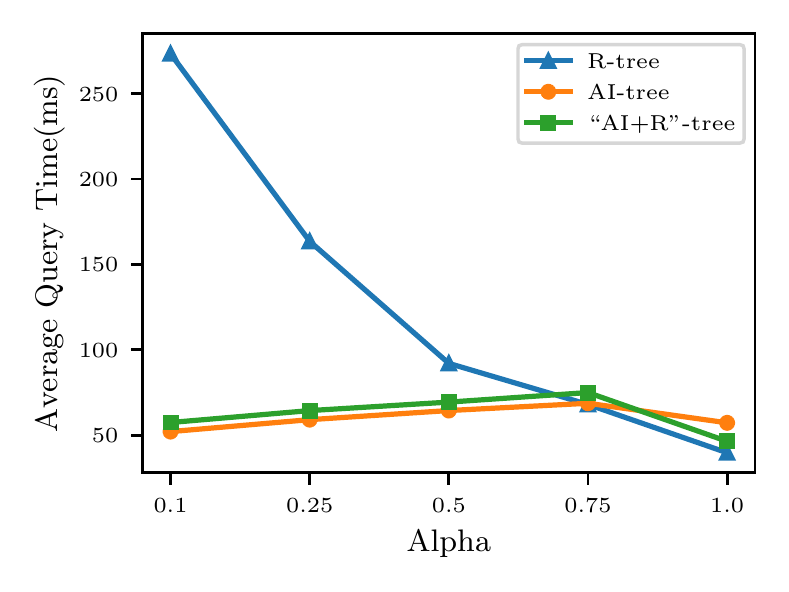}
        \caption{R-tree (M=400, m=200), and query selectivity = $0.00001$}
        \label{tw_ucr_me400_sel0.00001_10X10}
     \end{subfigure}
     \begin{subfigure}[b]{0.24\textwidth}
         \centering
       \includegraphics[width=\linewidth]{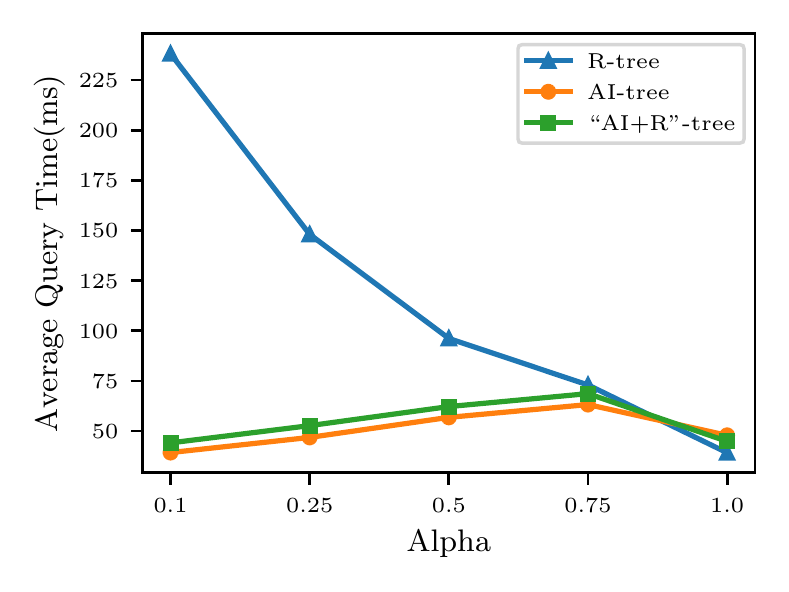}
        \caption{R-tree (M=800, m=400), and query selectivity = $0.00001$}
        \label{tw_ucr_me800_sel0.00001_10X10}
     \end{subfigure}
        \caption{Results on Tweet locations dataset} 
        \label{results_tweet_locations}
\end{figure*}

\subsubsection{Tweet Locations Dataset}
We  construct an R-tree using the minimum leaf capacity $m = 100$ and maximum leaf capacity $M = 200$. 
The selectivities of the synthesized
queries iare: $0.00001$, and $0.00005$. As a result, for each range query, the result contains approximately $20$ and $100$ data points, respectively. 
Moreover, the value of the threshold $\tau$ is set to $0.75$. 
In each of the figures for this dataset, we show the value of overlap ratio $\alpha$ in the X-axis and the average query processing time (in milliseconds) in the Y-axis. Also, we report the average query processing time taken by the 
standard 
R-tree, the AI-tree, and the ``AI+R''-tree. 




\subsubsection{Effect of Selectivity for the Tweets Location Dataset}



Figure~\ref{tw_ucr_me200_sel0.00001_20X20} gives the results for Selectivity $0.00001$.
From the figure, both the AI-tree and the ``AI+R''-tree enhance the performance of the R-tree by up to $3.69$X and $3.58$X, respectively.  Notice that the performance loss is minimal between the AI-tree and the ``AI+R''-tree, where the latter exhibits a hybrid approach to indexing. 
 This same pattern of performance gains for both trees applies for the cases of $\alpha = 0.25$. For $\alpha = 0.25$, the AI-tree and the ``AI+R''-tree enhance the performance of the R-tree up to $2.06$X and $1.97$X, respectively. Moreover, the ``AI+R''-tree performs better than the R-tree up to $\alpha = 0.50$. After that the R-tree starts to perform better. Notice that the hybrid approach reduces the query processing time of the AI-tree when the $\alpha = 1$.
 {\bf In summary, the ``AI+R''-tree gets the best of both worlds. In the case of high-overlap (low $\alpha$ value), the  ``AI+R''-tree performs similar to the AI-tree, while in the case of low-overlap (high $\alpha$ value), the ``AI+R''-tree
performs similar to the standard R-tree.}

Figure~\ref{tw_ucr_me200_sel0.00005_20X20} gives the results for the same setup but for a selectivity of  $0.00005$. As a result, each of the queries 
returns
approximately $100$ points.
From the figure, 
the AI-tree and the ``AI+R''-tree exhibit the similar trend in performance gains.

\subsubsection{The Effect of Node Capacity for the Tweets Location Dataset}
In the next experiment,
we vary the leaf node capacity. 
We cover the cases for $M=200, 400$, and $800$. 
We fix the selectivity of the 
synthesized queries to: $0.00001$ (The query result will contain approximately $20$ data points).
The AI-tree can perfectly fit the training data with a $10X10$ grid size for R-tree with node capacity $M = 400$, and $800$.

Figures~\ref{tw_ucr_me200_sel0.00001_20X20},~\ref{tw_ucr_me400_sel0.00001_10X10}, and~\ref{tw_ucr_me800_sel0.00001_10X10} give the performance results of the AI-tree and the ``AI+R''-tree for maximum leaf node capacities of 200, 400, and 800, respectively. 
The performance trends are the same. 
Overall, the performance gains of the AI-tree and the ``AI+R''-tree over the R-tree increase as the node capacities  increase. The reason is that as the node capacity increases, any additional extraneous leaf nodes retrieved by the traditional R-tree will be very expensive due to the refinement step. Basically, as the node capacity increases, more leaf data objects will need to be checked against the query range to refine the results and form the actual output data objects from among the ones in the leaf node.
In other words, due to the higher leaf node capacities, the penalty of an unnecessary scan inside an extraneous leaf node reduces the R-tree performance in contrast to the AI-tree and the ``AI+R''-tree.
In Figure~\ref{tw_ucr_me800_sel0.00001_10X10}, for node capacity 800, 
the AI-tree enhances the performance of the R-Tree up to $6.06$X for $\alpha = 0.10$. Also, the ``AI+R''-tree does not decrease the performance of the AI-tree by a large margin. To be precise, the ``AI+R''-tree  enhances the performance of the R-tree up to $5.39$X for $\alpha = 0.10$.

\begin{figure*}[tbph] 
     \centering
     \begin{subfigure}[b]{0.24\textwidth}
         \centering
        \includegraphics[width=\linewidth]{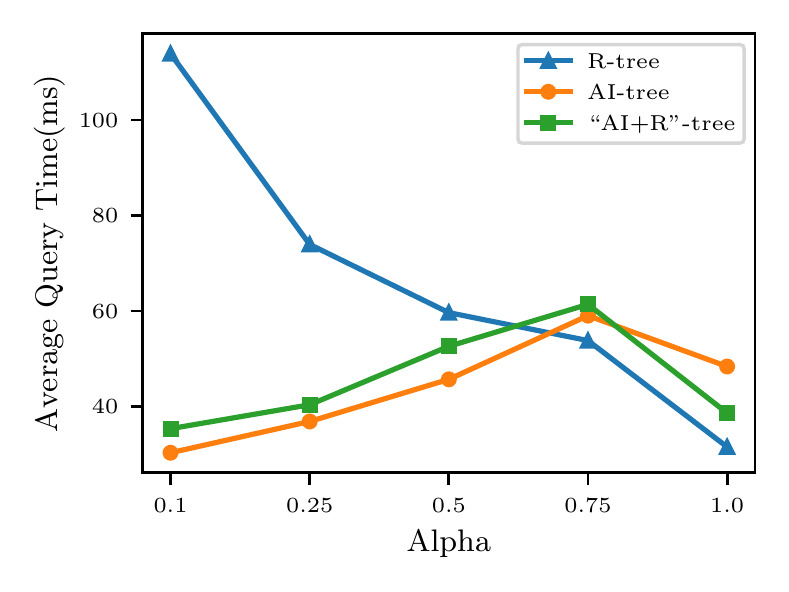}
        \caption{R-tree (M=200, m=100), and query selectivity = $0.00001$}
        \label{cc_me200_sel0.00001_20X20}
     \end{subfigure}
     \hfill
     \begin{subfigure}[b]{0.24\textwidth}
         \centering
         \includegraphics[width=\linewidth]{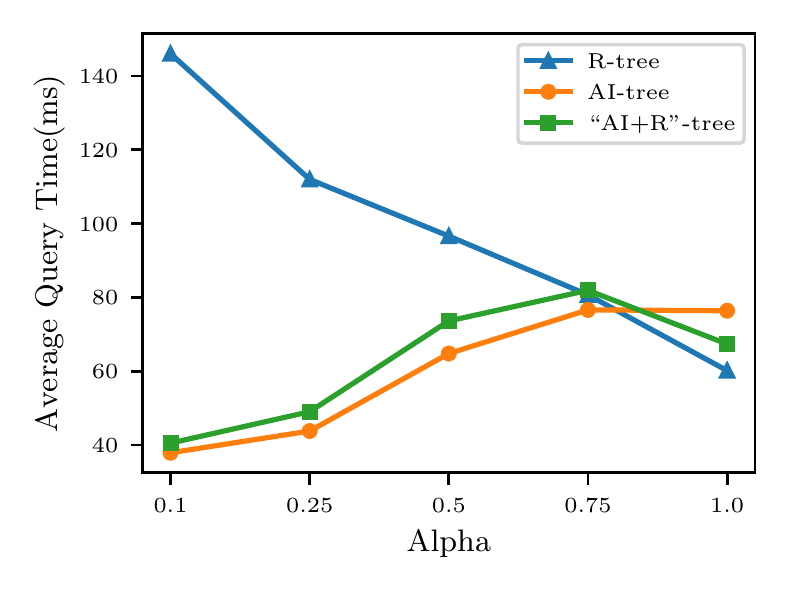}
        \caption{R-tree (M=200, m=100), and query selectivity = $0.00005$}
        \label{cc_me200_sel0.00005_20X20}
     \end{subfigure}
     \hfill
     \begin{subfigure}[b]{0.24\textwidth}
         \centering
        \includegraphics[width=\linewidth]{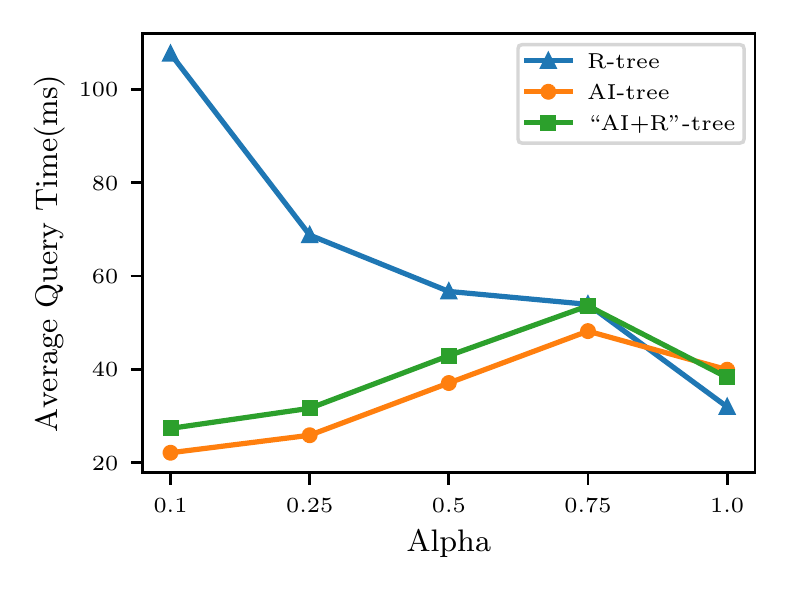}
        \caption{R-tree (M=400, m=200), and query selectivity = $0.00001$}
         \label{cc_me400_sel0.00001_10X10}
     \end{subfigure}
     \begin{subfigure}[b]{0.24\textwidth}
         \centering
        \includegraphics[width=\linewidth]{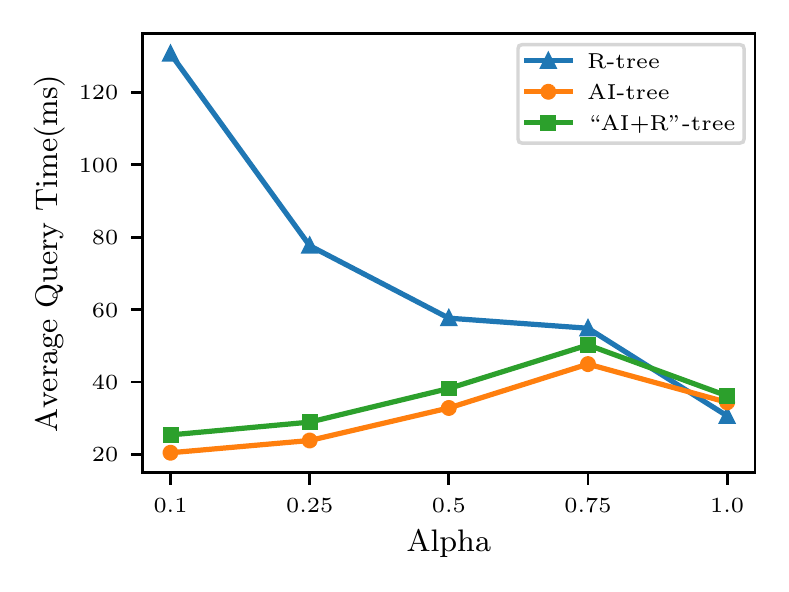}
        \caption{R-tree (M=800, m=400), and query selectivity = $0.00001$}
        \label{cc_me800_sel0.00001_10X10}
     \end{subfigure}
        \caption{Results on Chicago crimes dataset} 
        \label{results_chicago_crimes}
\end{figure*}

\subsubsection{Space Consumption of the ML Models for Tweets Location Dataset}
\begin{table}[h!]
    \caption{The R-tree and ML model sizes for the  ``AI+R''-tree  for each $\alpha$ (in MBs) for the Tweets Location dataset}
    \label{tab:modelsize_tweets_M=200_0.00001}
    \begin{tabular}{p{3em} p{3em} c c c c c c}
    \toprule
        &  &  &\multicolumn{5}{c}{``AI+R''-tree with various values of $\alpha$}\\
        \cmidrule(lr){4-8}
        Selec-tivity & Max Entries & R-tree & $0.10$ & $0.25$ & $0.50$ & $0.75$ & $1.0$\\
    \midrule
        0.00001 & 200 & 978.05 & 9.50 & 9.50 & 9.51 & 11.38 & 9.50\\
        0.00005 & 200 & 978.05 & 9.44 & 9.50 & 9.52 & 9.52 & 9.51\\
        0.00001 & 400 & 972.05 & 1.97 & 2.02 & 2.97 & 2.96 & 2.01\\
        0.00001 & 800 & 969.52 & 1.02 & 2.07 & 1.55 & 1.07 & 1.06\\
    \bottomrule
\end{tabular}
\end{table}
Table~\ref{tab:modelsize_tweets_M=200_0.00001} lists the sizes of the R-tree and the ML models of the ``AI+R''-Tree in Megabytes (MB, for short). Notice that the reported ML model size includes the sizes of both the multi-label and the binary classifiers.
The space requirements of the ML models for the ``AI+R''-tree with larger leaf capacity (e.g., M=400, and M=800) are less than those for the R-tree with leaf capacity $200$. 
The reason is that the number of leaf nodes is less for the larger node capacities. Also, 
notice that a grid of size $10X10$ is sufficient for ``AI+R''-tree with larger node capacity. Hence, less  models are likely to be trained to fit the data. Thus, the size of the ML models is even less than the cases of using a larger grid of size $20X20$.


\subsubsection{The Chicago Crimes Dataset}
Overall, the Chicago Crimes dataset reflects the same performance trends as those for the Tweets Location dataset in favor of  the ``AI+R''-tree over the R-tree. We give the  performance results for the the Chicago Crimes dataset below.
For the Chicago crimes dataset, initially, we construct an R-tree using the maximum leaf capacity $M = 200$, and  minimum leaf capacity $m = 100$. Moreover, the selectivity of the synthesized
queries is: $0.00001$ and $0.00005$. For each range query, the result contains (approximately) $9$ and $44$ data points, respectively. Moreover, the AI-tree can perfectly fit the training data for this query workload with a grid size $20X20$.  $\tau$ is set to $0.75$. 

\subsubsection{Effect of Selectivity for the Chicago Crimes Dataset}



Figures~\ref{cc_me200_sel0.00001_20X20} and~\ref{cc_me200_sel0.00005_20X20} give the performance results for the AI-tree, the ``AI+R''-tree, and the R-tree for Selectivities 0.00001 and 0.00005.
The performance gains of the ``AI+R'' over the R-tree is up to $3.6$X for high overlap queries ($\alpha = 0.10$) while is very close to the R-tree for $\alpha = 1.0$. This is consistent for both selectivity values.

\subsubsection{Effect of Node Capacity for Chicago Crimes Dataset}
We vary leaf node capacity to cover for $M=200, 400$, and $800$. 
We fix query selectivity to: $0.00001$.
%
Moreover, the AI-tree can perfectly fit the training data for a $10X10$ grid size and R-tree with node capacity $M = 400$ and $800$.
Figures~\ref{cc_me200_sel0.00001_20X20},~\ref{cc_me400_sel0.00001_10X10}, and~\ref{cc_me800_sel0.00001_10X10} give the performance results of the AI-tree and the ``AI+R''-tree for the Chicago Crimes dataset for maximum leaf node capacities $M=$ 200, 400, and 800. 
The performance trends are consistent with those of the Tweets Location dataset. 
Overall, the performance gains of the AI-tree and the ``AI+R''-tree over the R-tree increase as the node capacities  increase, e.g., for $M=800$, the ``AI+R''-tree enhances the performance over the R-tree by $5.14$X for $\alpha = 0.10$. Also, the ``AI+R''-tree reduces the query performance of the AI-tree for $\alpha = 1$ to be close to that of the R-tree.

\subsubsection{Space Consumption of the ML Models for the Chicago Crimes Dataset}

Table~\ref{tab:modelsize_chicago_M=200_0.00001}
lists the sizes of the R-tree and the ML models of the ``AI+R''-Tree (in MBs). The reported ML model size 
includes the sizes of both the multi-label and the binary classifiers. 
The space requirements of the ML models for the ``AI+R''-tree with larger leaf capacity (e.g., M=400, and M=800) are less than for the R-tree with $M=200$. A grid of size $10X10$ is found to be sufficient for the ``AI+R''-tree in these cases. Hence, less models are likely to be trained to fit the data. Also, the size of the ML models are even less than the cases for using a larger grid of size $20X20$.

\begin{table}[h!]
    \caption{The R-tree and  ML model sizes for the  ``AI+R''-tree for each $\alpha$ (in MBs) for the Chicago Crimes dataset.} 
    \label{tab:modelsize_chicago_M=200_0.00001}
    \begin{tabular}{p{3em} p{3em} c c c c c c}
    \toprule
        &  &  &\multicolumn{5}{c}{``AI+R''-tree with various values of $\alpha$}\\
        \cmidrule(lr){4-8}
        Selec-tivity&Max Entries & R-tree&$0.10$&$0.25$&$0.50$&$0.75$&$1.0$\\
    \midrule
        0.00001 & 200 & 426.46 & 3.39 & 2.57 & 3.40 & 3.40 & 3.40\\
        0.00005 & 200 & 426.46 & 2.52 & 2.57 & 2.58 & 3.41 & 3.41\\
        0.00001 & 400 & 423.86 & 0.48 & 0.53 & 0.53 & 0.53 & 0.53\\
        0.00001 & 800 & 422.84 & 0.27 & 0.32 & 0.33 & 0.33 & 0.32\\
    \bottomrule
\end{tabular}
\end{table}

\subsubsection{\textbf{Discussion}}



From Figures~\ref{tw_ucr_me200_sel0.00001_20X20}  and~\ref{cc_me200_sel0.00001_20X20}, the R-tree performs better than the AI-tree for $\alpha = 0.75$. As we set the threshold $\tau = 0.75$, the ``AI+R''-tree also degrades in performance because it uses the AI-tree to process these queries with $\alpha = 0.75$. In both  cases, the R-tree has relatively small leaf capacity (i.e., $M=200$). As the leaf capacity increases, for the same value of threshold $\tau=0.75$, the ``AI+R''-tree performance enhances (see Figure~\ref{tw_ucr_me800_sel0.00001_10X10} for the Tweets Location dataset and Figure~\ref{cc_me800_sel0.00001_10X10} for Chicago Crimes dataset).

For each $\alpha$,  the ML models increase the space requirement of the R-tree by no more than $1.1\%$ (see Table~\ref{tab:modelsize_tweets_M=200_0.00001}). Also, {\bf the space overhead of the ML models for all values of $\alpha$ does not increase the size of the R-tree by more than $5.04\%$} (Table~\ref{tab:modelsize_tweets_M=200_0.00001}). Thus, the space requirement of the ML models can be as low as $0.37\%$ of the R-tree size (Table~\ref{tab:modelsize_chicago_M=200_0.00001}).

\section{Related Work} \label{section:related_work}
Many variants of the R-tree have been introduced, 
e.g., see~\cite{guttman1984r, sellis1987r+, beckmann1990r, beckmann2009revised, samet2006foundations, manolopoulos2010r}.
The R$^{+}$-tree~\cite{sellis1987r+} 
recognizes the problem of node overlap in the R-tree, and creates an R-tree so that no two nodes overlap in space. 
The R$^{*}$-tree~\cite{beckmann1990r} reduces node overlap by introducing the forced re-insertion of entries.
The RR$^{*}$-tree~\cite{beckmann2009revised} is a 
further improvement over
the
R$^{*}$-tree for dynamic data. 
The RR$^{*}$-tree improves 
over 
the R$^{*}$-tree by restricting the insertion to a single path and dropping the idea of re-insertion.
In~\cite{sidlauskas2018improving}, the Clipped Bounding Box (CBB) based R-tree further improves the I/O performance of the R$^{*}$-tree.
The priority R-tree~\cite{arge2008priority}
can answer a query with an asymptotically optimal number of I/Os. 
The Hilbert R-tree~\cite{kamel1993hilbert} leverages the Hilbert space-filling curve to impose an ordering on the R-tree nodes to achieve good space utilization.
A worst-case optimal R-tree packing strategy that uses space-filling curves can be found in~\cite{qi2020packing}. 
Notice that regardless of the type of the R-tree, all R-tree variants attempt to reduce the amount of node overlap. However, with dynamic updates, the shape of an constructed R-tree deteriorates. 
As a result, our design principles for ``AI+R''-tree can be applied to other R-tree variants. 

The concept of separating objects into partitions based on their size and indexing each partition with a space filling curve can be found in ~\cite{zhang2014towards, koudas1997size}. However, in the case of ``AI+R''-tree, we do not partition the objects based on their size, but rather we group the queries using the grid to train multiple ML models. Moreover, we do not use any space filling curve (i.e., as a projection function).

The initial research on learned indexes ~\cite{kraska2018case, CDFShop} has introduced the idea that ``Indexes are models'' by proposing a Recursive Model Index (RMI, for short) for read-only workloads.
Many followup research has been conducted that is inspired by RMI both in the single and Multi-dimensional space \cite{ferragina_survey, al2020tutorial,idreos2019_tutorial,sabek2020machine_tutorial}.
In the case of multi-dimensional indexes, some initial effort to extend the idea of RMI into the multi-dimensional space can be found in~\cite{kraska2019sagedb}. In~\cite{wang2019learned}, Z/Morton order is used to project the data into the one dimensional space. Then, an RMI-like structure can be used to build the learned index.
However, learning the projection function, e.g., Z-order~\cite{sagan2012space, mokbel2003analysis}, from the multi-dimensional space to the one-dimensional space is hard. Thus, it has been proposed to choose a layout that is easy to learn by 
an
ML model. 
An efficient scaling method has been proposed in~\cite{davitkova2020ml}.
In our proposed ``AI+R''-tree, we avoid using a projection function, and operate directly on the original multi-dimensional representation of the spatial data objects. 
In~\cite{nathan2019learning}, an in-memory learned multi-dimensional index, termed Flood, is introduced to efficiently support queries for a particular dataset and (read-only ) query workloads. An extension to Flood has been proposed that can adapt to changes in the query workload~\cite{ding2020tsunami}.
Reinforcement Learning has been used to build an efficient data layout~\cite{yang2020qd, ding2021instance} for a particular dataset and query workload. A learned spatial index for disk-based systems can be found in~\cite{LISA}. In~\cite{qi13effectively}, another disk-based spatial index, termed RSMI, leverages a rank-space-based transformation. The transformation has been used to get an easily learnable CDF.
Notice that the goal of the above mentioned learned multi-dimensional indexes is to replace a traditional index. However, in the case of the ``AI+R''-tree, our target is not to replace the existing index structure rather to enhance its performance using ML models. 

The idea of using helper ML models inside traditional indexes to enhance their performance have been presented 
in the multi-dimensional space, e.g., see~\cite{hadian2020handsoff, pandey2020case, kang2021the}. In~\cite{hadian2020handsoff}, interpolation-based learned spatial indexes 
are proposed. In~\cite{pandey2020case}, techniques from~\cite{nathan2019learning} have been applied to five traditional multi-dimensional indexes.
Recently, a disk-based ML-enhanced index to process k-nearest-neighbor queries over high-dimensional time-series data has been proposed~\cite{kang2021the}. The goal of the proposed method~\cite{kang2021the} is to re-organize the access order of the leaf nodes. 
In the context of ML-enhanced multi-dimensional indexes, the focus of the above mentioned techniques is not on analyzing (i.e., high- vs. low-overlap queries) and optimizing the index for a given
query workload. 
Notice that, in the case of the ``AI+R''-tree, the focus is on analyzing the query workload to identify the queries for which a traditional disk-based spatial index (in this case, the R-tree) does not perform well.
Moreover, we propose to adopt a hybrid approach to leverage the benefit of both the proposed AI-tree, and the traditional R-tree.

Surveys on the topic of learned data structures can be found in~\cite{ferragina_survey, zhou2020database}. Recently, several tutorials related to learned indexes have been presented in different venues~\cite{li2021ai_tutorial,idreos2019_tutorial,sabek2020machine_tutorial,al2020tutorial,echihabi2021new}.

\section{Conclusion}\label{section:conclusion}
In this paper, 
we leverage machine learning techniques to 
build an instance-optimized R-tree for a given data and query workloads. 
Although the paper focuses on the R-tree, the proposed design principles in the paper apply to other spatial indexes as long as node overlaps exist, and hence multiple tree paths are explored during search. Notice that we avoid using a projection function, and operate directly on the original representations of the spatial data objects.
Also, because the ``AI+R''-tree operates at the leaf node level of an R-tree, the proposed method can support different types of objects (e.g., objects with extension). 
Additionally, we adopt a multi-model approach and index the learned ML models using a grid-based structure. We further leverage ML techniques to train a binary classifier to differentiate between high- and low-overlap queries. Finally, we advocate for a hybrid approach, namely the ``AI+R''-tree by combining both the traditional R-tree structure and the learned R-tree (i.e., the AI-tree) structure to maximize query processing performance.
In the future, we plan to investigate alternative choices for the ML models, and how to support k-NN query and spatial join using the proposed ``AI+R''-tree. 
As we maintain a hybrid structure inside  the ``AI+R''-tree, we will be able to sustain updates using its R-tree component. 
However, propagating the updates to the AI-tree component is an interesting future research direction.
Finally, we plan to investigate challenges related to the  integration of the proposed ``AI+R''-tree into practical database systems.


\section*{Acknowledgments}
Walid Aref acknowledges the support of the U.S. National Science Foundation under Grant Numbers: III-1815796 and  IIS-1910216.


\balance
\bibliographystyle{IEEEtran}
\bibliography{reference}

\end{document}